\documentclass[a4paper,11pt]{article}
\usepackage{pos}
\newcommand{\Bern}{Institute for Theoretical Physics, Albert Einstein Center for Fundamental Physics
, Switzerland
}
\newcommand{\hiskp}{HISKP (Theory), Rheinische Friedrich-Wilhelms-Universit\"at Bonn, Germany
}
\newcommand{\hpca}{High Performance Computing and Analytics Lab, Rheinische Friedrich-Wilhelms-Universit\"at Bonn, Germany}
\newcommand{\CyprusU}{Department of Physics, University of Cyprus, 20537 Nicosia, Cyprus}
\newcommand{\CyprusI}{Computation-based Science and Technology Research Center, The Cyprus Institute, Cyprus}
\newcommand{\Jena}{University of Jena, Institute for Theoretical Physics, Max-Wien-Platz 1, D-07743 Jena, Germany}
\newcommand{\Parma}{Dipartimento  di  Scienze  Matematiche,  Fisiche  e  Informatiche,  Universit\`a  di  Parma  and  INFN,  Italy}
\newcommand{\Romauno}{Dipartimento di Fisica and INFN, Universit\`a di Roma ``La Sapienza", I-00185 Rome, Italy}
\newcommand{\Romadue}{Dipartimento di Fisica and INFN, Universit\`a di Roma ``Tor Vergata", I-00133 Rome, Italy}
\newcommand{\Romatre}{Dipartimento di Matematica e Fisica, Universit\`a Roma Tre and INFN, Sezione di Roma Tre, Italy}
\newcommand{\RomatreINFN}{Istituto Nazionale di Fisica Nucleare, Sezione di Roma Tre, Italy}
\newcommand{\temple}{Department of Physics,  Temple University,  Philadelphia, PA 19122 - 1801, USA}
\newcommand{\RomaunoINFN}{Istituto Nazionale di Fisica Nucleare, Sezione di Roma La Sapienza, Italy}
\newcommand{\Edin}{
School of Physics and Astronomy,
The University of Edinburgh, Edinburgh EH9 3FD, UK}
\newcommand{\Odense}{CP$^3$-Origins, University of Southern Denmark, Campusvej 55, 5230 Odense, Denmark}
\newcommand{\Grenoble}{Theory Group, Laboratoire de Physique Subatomique et de Cosmologie, Grenoble, France}
\newcommand{\Fermi}{Centro Fermi, Museo Storico della Fisica e Centro Studi e Ricerche ``Enrico Fermi", Italy}
\newcommand{\NIC}{NIC, DESY, Platanenallee 6, D-15738 Zeuthen, Germany}
\newcommand{\Berlin}{Institut f\"ur Physik, Humboldt-Universit\"at zu Berlin, Newtonstrasse 15, 12489 Berlin, Germany}

\title{Determination of the light, strange and charm quark masses using twisted mass fermions}

\author*[a,b]{C.~Alexandrou}\affiliation[a]{\CyprusU}\affiliation[b]{\CyprusI}
\author[b]{S.~Bacchio}
\author[c]{G.~Bergner}\affiliation[c]{\Jena}
\author[d]{M.~Constantinou}\affiliation[d]{\temple}
\author[e,f]{M.~Di Carlo}\affiliation[e]{\Edin}\affiliation[f]{\RomaunoINFN}
\author[g]{P.~Dimopoulos}\affiliation[g]{\Parma}
\author[b]{J.~Finkenrath}
\author[h]{E.~Fiorenza}\affiliation[h]{\Odense}
\author[i]{R.~Frezzotti}\affiliation[i]{\Romadue} 
\author[j]{M.~Garofalo}\affiliation[j]{\hiskp}
\author[a,b]{K.~Hadjiyiannakou}
\author[k]{B.~Kostrzewa}\affiliation[k]{\hpca}
\author[b]{G.~Koutsou}
\author[l]{K.~Jansen}\affiliation[l]{\NIC}
\author[m]{V.~Lubicz}\affiliation[m]{\Romatre}
\author[n]{M.~Mangin-Brinet}\affiliation[n]{\Grenoble}
\author[a,i,o]{F.~Manigrasso}\affiliation[o]{\Berlin}
\author[p]{G.~Martinelli}\affiliation[p]{\Romauno}
\author[b]{F.~Pittler}
\author[i,q]{G.C.~Rossi}\affiliation[q]{\Fermi}
\author[r]{F.~Sanfilippo}\affiliation[r]{\RomatreINFN}
\author[r]{S.~Simula}
\author[m]{C.~Tarantino}
\author[a,i,o]{A.~Todaro}
\author[j]{C.~Urbach}
\author[s]{U.~Wenger}\affiliation[s]{\Bern}

\abstract{We present results for the light, strange and charm quark masses using $N_f=2+1+1$ twisted mass fermion ensembles at three values of the lattice spacing, including two ensembles simulated with the physical value of the pion mass. The analysis is done both in the meson and baryon sectors. The difference in the mean values found in the two sectors is included as part of the systematic error. The presentation is based on the work of Ref.~\cite{Alexandrou:2021gqw}, where more details can be found. }

\FullConference{%
 The 38th International Symposium on Lattice Field Theory, LATTICE2021
  26th-30th July, 2021
  Zoom/Gather@Massachusetts Institute of Technology
}

\begin{document}
\maketitle

\section{Introduction}
Quark masses are crucial inputs  for the phenomenological description of the plethora of  phenomena governed by the strong nuclear force. We  use several ensembles simulated using the twisted mass fermion action at  three values of the lattice spacing and spanning pion masses in the range from about 350 MeV to 135 MeV. This enable us to perform a combined chiral and continuum extrapolation. 
The  properties of the ensembles used in this work are summarized in Table~\ref{tab:params}. We will refer to the ensembles in Table~\ref{tab:params} with the names starting with $cA$ in their names as A ensembles, those starting with $cB$ as B ensembles and those with $cC$ as C ensembles. 

In order to avoid undesired ${\cal O}(a^2)$ mixing of the strange and charm flavours in our
physical observables, we adopt a non-unitary lattice setup~\cite{Frezzotti:2004wz}, where the twisted-mass action for non-degenerate strange and charm quarks is employed only in the sea sector, while the valence strange and charm quarks that enter the correlation functions are regularized as exactly flavour-diagonal Osterwalder-Seiler fermions~\cite{Osterwalder:1977pc}.

\begin{table}[htb!]
	\centering
	\small
	\setlength{\tabcolsep}{5pt}
	\begin{tabular}{|lccccccr|}
		\hline
		Ensemble       &  $L^3\times T$  & $~a\mu_\ell$ &   $am_\pi $    &   $af_\pi$    & $~m_\pi L~$ &   $m_N/m_\pi$  & $m_\pi$ [MeV] \\ \hline
		\multicolumn{8}{c}{$\beta=1.726$, $c_{SW}=1.74$, $a\mu_\sigma = 0.1408$, $a\mu_\delta=0.1521$, $w_0/a = 1.8352~(35)$}    \\ \hline
		cA211.53.24  & $24^3\times48$  &   0.00530   & 0.16626 (51)  & 0.07106 (36) &   3.99    &          --      & 346.4 (1.6) \\
		cA211.40.24  & $24^3\times48$  &   0.00400   & 0.14477 (70)  & 0.06809 (30) &   3.47    &          --      & 301.6 (2.1) \\
		cA211.30.32  & $32^3\times64$ &   0.00300  & 0.12530 (16) & 0.06674 (15) &   4.01    &   4.049 (14) & 261.1 (1.1) \\
		cA211.12.48  & $48^3\times96$  &   0.00120   & 0.08022 (18)& 0.06133 (33) &   3.85    &   5.685 (28) & 167.1 (0.8) \\ \hline
		\multicolumn{8}{c}{$\beta=1.778$, $c_{SW}=1.69$, $a\mu_\sigma = 0.1246864$, $a\mu_\delta=0.1315052$, $w_0/a = 2.1299~(16)$ } \\ \hline
		cB211.25.32  & $32^3\times64$  &   0.00250   &  0.10475 (45) & 0.05652 (38) &   3.35    &   4.104 (36) & 253.3 (1.4) \\
		cB211.25.48  & $48^3\times96$  &   0.00250   & 0.10465 (14) &  0.05726 (12) &   5.02    &   4.124 (17) & 253.0 (1.0) \\
		cB211.14.64  & $64^3\times128$  &   0.00140   & 0.07848 (10) &  0.05477 (12) &   5.02    &   5.119 (36) & 189.8 (0.7) \\
		cB211.072.64 & $64^3\times128$  &   0.00072   & 0.05659 (8) &  0.05267 (14) &   3.62    &   6.760 (30) & 136.8 (0.6) \\ \hline
		\multicolumn{8}{c}{$\beta=1.836$, $c_{SW}=1.6452$, $a\mu_\sigma = 0.106586$, $a\mu_\delta=0.107146$, $w_0/a = 2.5045~(17)$}  \\ \hline
		cC211.20.48  & $48^3\times96$  &   0.00200   & 0.08540 (17) & 0.04892 (13) &   4.13    &   4.244 (25) & 245.73 (98)\\
		cC211.06.80  & $80^3\times160$  &   0.00060   & 0.04720 (7) & 0.04504 (10) &   3.78    &   6.916 (19) & 134.3 (0.5) \\ \hline
	\end{tabular}
	\caption{Parameters of the $N_f=2+1+1$ ensembles analyzed in this study. In the first column we give the name of the ensemble, in the second the lattice volume, in the third the twisted-mass parameter, $a\mu_\ell$, for the average up/down (light) quark, in the fourth and in the fifth the pion mass $am_\pi$ and decay constant $af_\pi$ in lattice units from Ref.~\cite{Alexandrou:2021bfr}, in the sixth the pion mass times the lattice spatial length, $m_\pi L$, in the seventh the ratio $m_N / m_\pi$ as determined in Section~\ref{sec:baryons} and, finally, in the last column the pion mass in physical units, using our determination of the gradient-flow scale $w_0$ obtained in Ref.~\cite{Alexandrou:2021bfr}. We also include for each set of ensembles with the same lattice spacing the coupling constant $\beta$, the clover-term parameter $c_{SW}$, the parameters of the non-degenerate operator $a\mu_\sigma$ and $a\mu_\delta$, related to the renormalized strange and charm sea quark masses~\cite{Frezzotti:2004wz}, and the value of the gradient-flow scale $w_0/a$ determined at the physical pion mass in Ref.~\cite{Alexandrou:2021bfr}.
	}
	\label{tab:params}
\end{table}

A new feature of this work is the use of two sets of observables to set the scale and to evaluate the quark masses enabling us to  study systematic effects in the determination of the quark masses using different inputs. One set of observables is based on quantities from the meson sector, namely we use the pion mass and decay constant to set the scale and to determine the average up/down quark mass, referred thereafter as light quark mass,  and the kaon and $D$-meson masses for the determination of the mass of the strange and charm quarks, respectively. In the baryon sector, the nucleon and pion masses are used  to set the scale and light  quark mass and  the $\Omega^-$ and the $\Lambda_c$ masses are used to determine the strange and charm quark masses, respectively. In our analysis in the baryon sector we restricted ourselves to using gauge ensembles simulated  with pion masses less than 300~MeV. 

Another feature of this work is the improved determination of the renormalization factor $Z_P$. In the maximally twisted-mass formulation used here the  renormalized quark mass is given  by $m_q = \mu_q/ Z_P$ and it is, thus, a crucial input for determining the quark masses. Our approach to compute $Z_P$ is described in more detail in Ref.~\cite{Alexandrou:2021gqw} and presented at this conference in Ref.~\cite{DiCarlo}.

  \section{Determination of the lattice spacing}
\noindent
{\it Meson sector}.  We use the iso-symmetric values of the pion  mass and decay constant, given respectively by~\cite{Aoki:2016frl},
\begin{equation}\label{eq:pi_isoqcd}
    m^{isoQCD}_\pi=~135.0(2)\text{~MeV}\quad\text{and}\quad f^{isoQCD}_\pi =~130.4 (2)\text{~MeV} ~ .
\end{equation}
NLO SU(2) chiral perturbation theory is employed, to correct for volume effects and take the continuum limit of $m_\pi$  and $f_\pi$  in units of $w_0$.
 Using  $w_0/a$ computed for for each gauge ensemble  we extrapolate to the physical pion mass and continuum limit. We find $w_0 =~0.17383(63)$ fm~\cite{Alexandrou:2021bfr} and using this value we determine   the three lattice spacings. Details are given in Ref.~\cite{Alexandrou:2021bfr}.\\
 {\it Baryon sector.}
 We use the iso-symmetric values of the pion and nucleon mass, $m_N^{\rm isoQCD}$=~0.938~GeV and SU(2) chiral perturbation theory to one-loop
 \begin{equation}
 (a_i m_N) = a_i m_N^0 - 4c_1 \frac{(a_im_\pi)^2}{a_i} - \frac{3g_A^2}{16\pi f_\pi^2} \frac{(a_im_\pi)^3}{a_i^2},
\label{eq:nucl_fit}
 \end{equation}
 where $a_i$ are the three lattice spacings, $m_N^0$ is the nucleon mass at the chiral limit and $c_1$ is fixed using the value of  $m_N^{\rm isoQCD}$.  The axial charge $g_A$ is set to its physical value of $g_A = 1.27641(56)$.
\begin{figure}[htb!]
\begin{minipage}{0.5\linewidth}
\includegraphics[width=\linewidth]{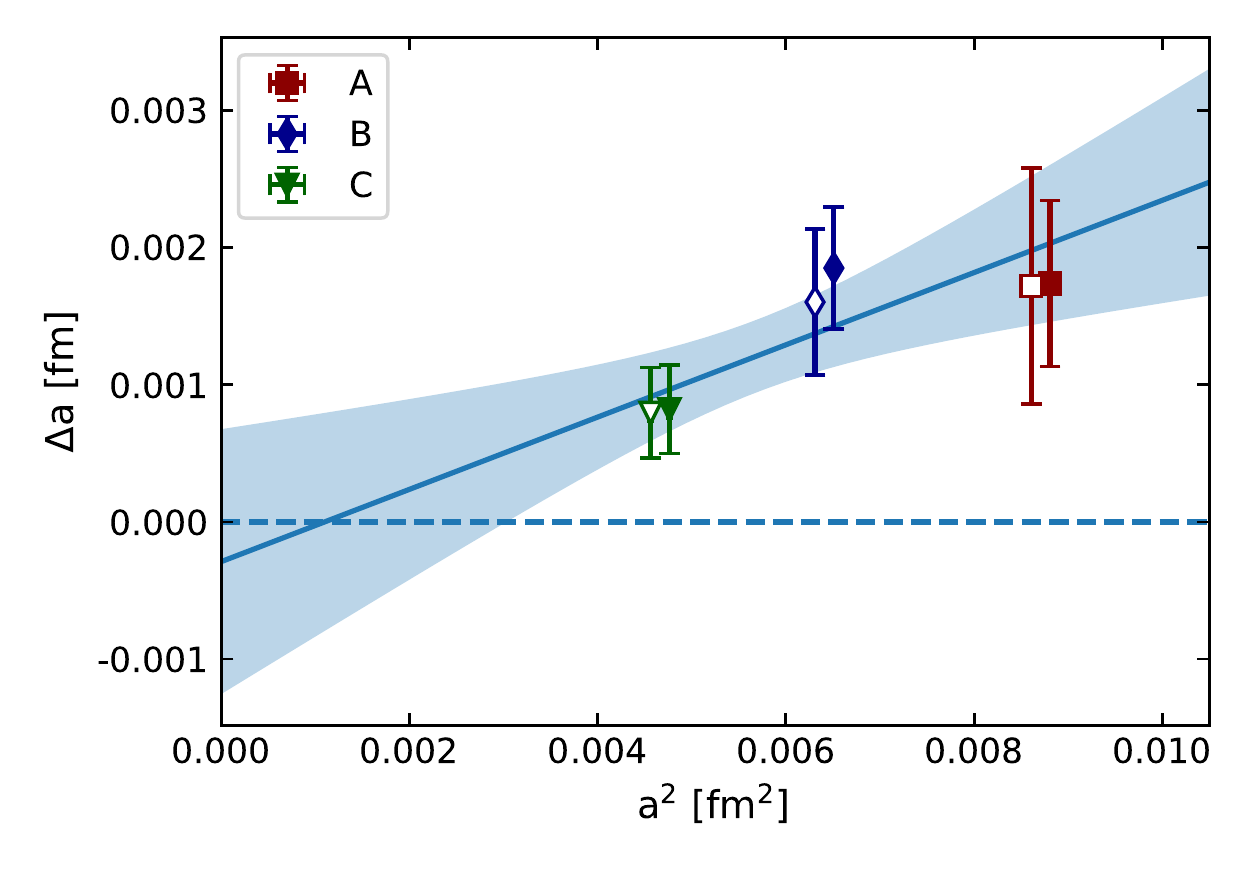}
   \vspace{-0.5cm}
\end{minipage}
\begin{minipage}{0.5\linewidth}
\small
\setlength{\tabcolsep}{4pt}
  \begin{tabular}{|l|ccc|}
\hline
Sector & $a_A$ [fm] & $a_B$ [fm] & $a_C$ [fm] \\\hline\hline
Pion &$0.09471(39)$ & $0.08161(30)$ & $0.06942(26)$\\\hline
Nucleon &$0.09295(47)$ & $0.07975(32)$ & $0.06860 (20)$\\\hline\hline
$\Delta a$ &$0.00176(61)$ & $0.00186(44)$ & $0.00082(32)$\\\hline
  \end{tabular}
\end{minipage}
   \caption{Left: The difference $\Delta a$ between the lattice spacings determined from the pion sector and the nucleon mass versus  $a^2$. Full symbols are the lattice spacings determined using all the ensembles for which $m_\pi<260$~MeV. Open symbols, shifted to the left for clarity, are obtained using ensembles for which the pion mass is below 190~MeV. The solid line shows the linear fit in $a^2$ to the results extracted by using ensembles with $m_\pi<260$~MeV (full symbols), which is largely consistent with zero in the continuum limit. Right: The values of the lattice spacings.}\label{fig:scale}
\end{figure}

The values of the lattice spacing extracted from the pion sector and from the nucleon mass  differ by ${\cal O}(a^2)$ effects. Fitting their difference as a function of $a^2$, as shown in Fig.~\ref{fig:scale}, we observe that in the continuum limit the difference vanishes, as expected.

\section{Determination of  quark masses in the meson sector}\label{sec:mesons}
To determine  the light quark mass, we use   SU(2) chiral perturbation theory (ChPT) for $m_\pi$ and $f_\pi$ given by

\begin{equation}
\label{eq:cptmpi2Ch}
(m_\pi  w_0 )^2   =  2(B w_0 )(m_\ell w_0 )\left[ 1 + \xi_\ell \log \xi_\ell  + P_1 \xi_\ell  +  P_2  \,a^2/w_0^2 \right] K_{M^2}^{FSE}
\end{equation}      
\begin{equation}
\label{eq:cptfpiCh}
(f_\pi  w_0 )  =  (f w_0 )\left[ 1 - 2\xi_\ell \log \xi_\ell  + P_3 \xi_\ell  +  P_4 \,a^2/w_0^2 + a^2m_\ell P_5 \right] K_f^{FSE},
\end{equation}      
where $\xi_\ell=\frac{2B m_\ell}{(4\pi f)^2}$, $P_1=-\bar\ell_3-2\log{\left(m_\pi^{\textrm{isoQCD}}/(4\pi f)\right)}\,,\quad
P_3=2\bar\ell_4+4\log{\left(m_\pi^{\textrm{isoQCD}}/(4\pi f)\right)}$ and the quantities $K_{M^2}^{FSE}$ and $K_f^{FSE}$ represent the finite size effects (FSE) on the squared pion mass and the pion decay constant, respectively.
\begin{figure}[htb!]
    \centering
    \begin{minipage}{0.5\linewidth}
    \includegraphics[width=1.0\textwidth]{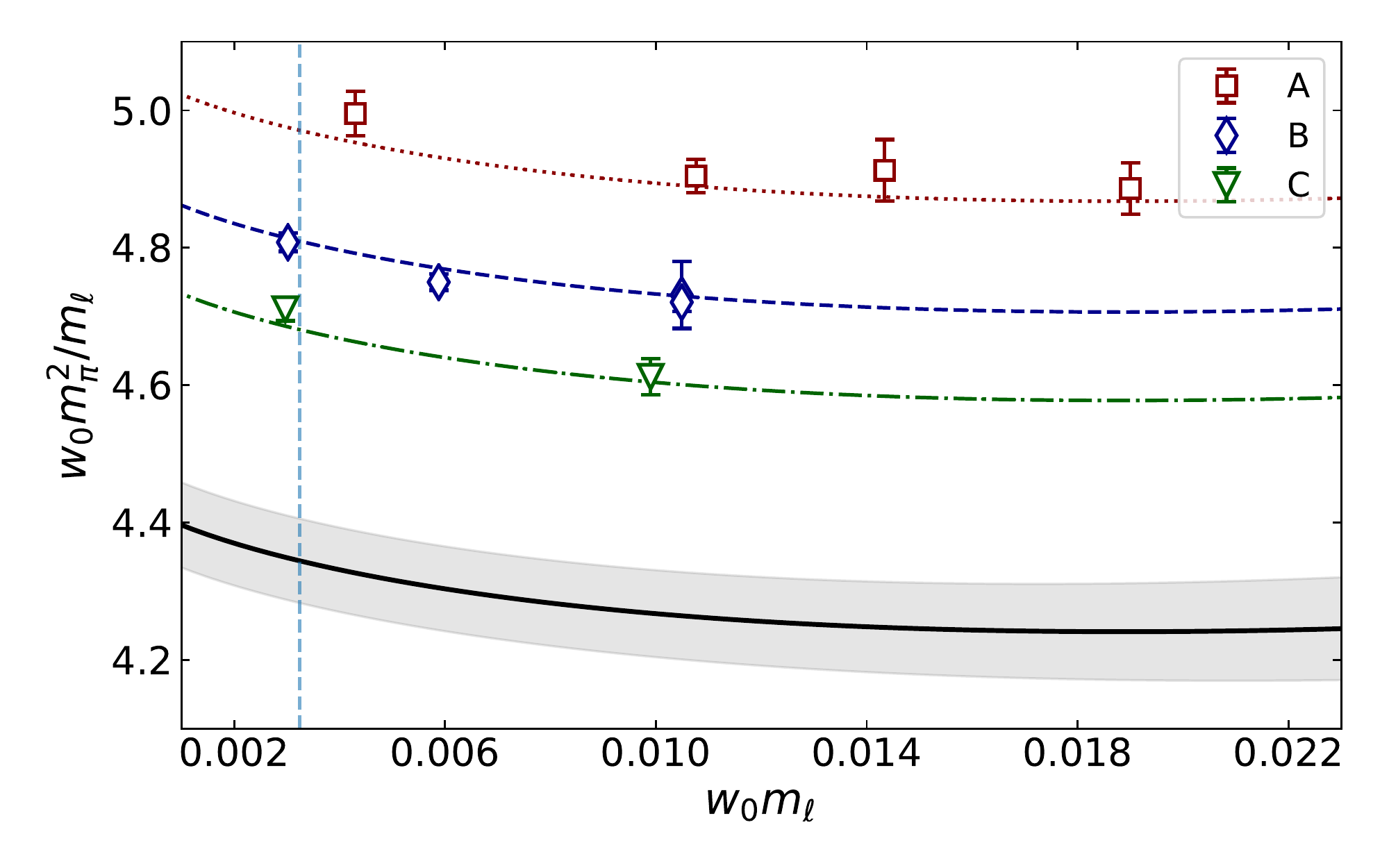}
    \end{minipage}%
    \begin{minipage}{0.5\linewidth}
    \includegraphics[width=1.0\textwidth]{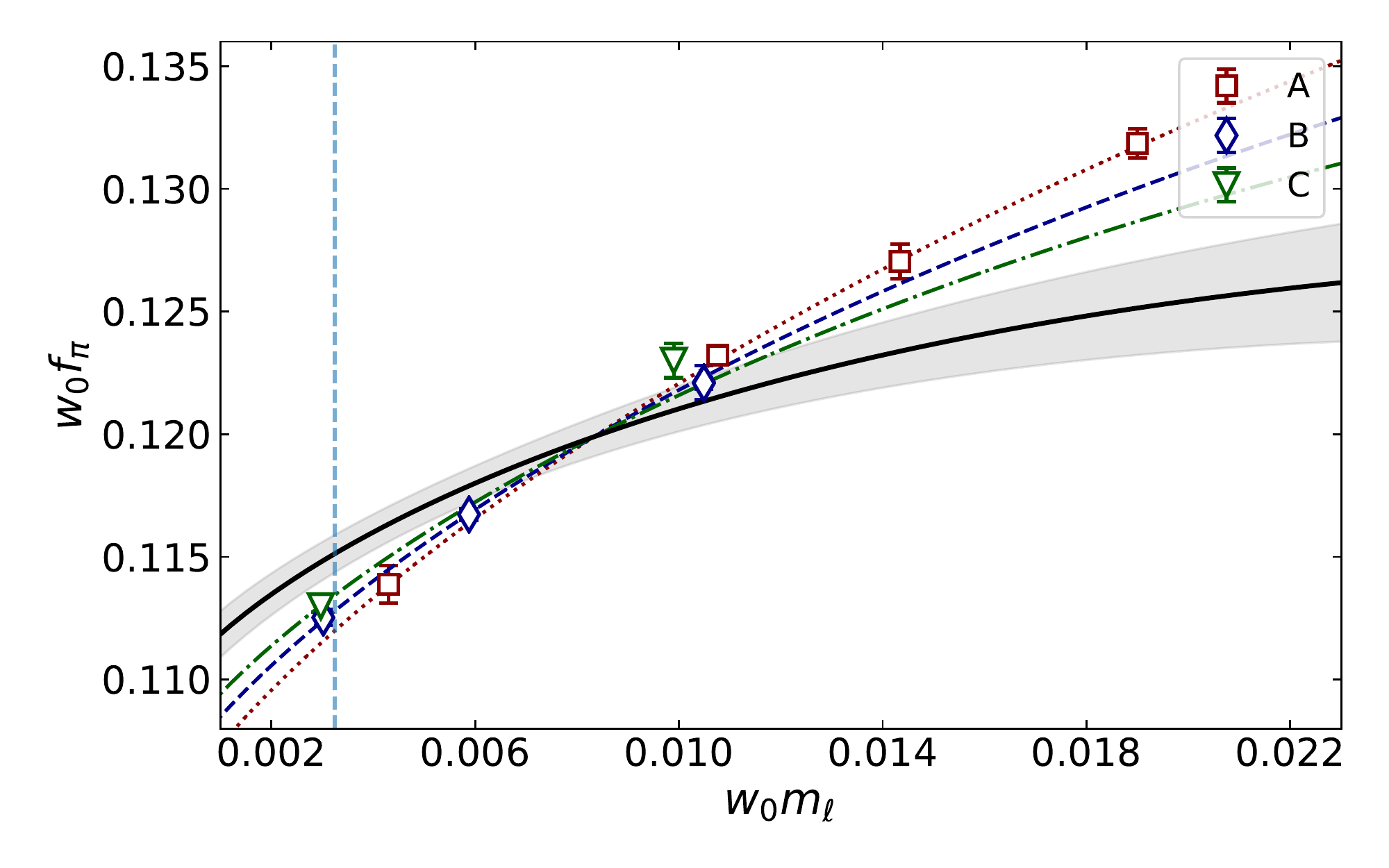}
    \end{minipage}
    \vspace{-0.5cm}
    \caption{Chiral and continuum extrapolation of $w_0 m_{\pi}^2/m_{\ell}$ (left) and $w_0 f_{\pi}$ (right) as function of $w_0\,m_\ell$ using Eqs.~(\ref{eq:cptmpi2Ch}) and (\ref{eq:cptfpiCh}) and $Z_P$ for the M2b method. Different colored bands correspond to different lattice spacings (red for the A ensembles, blue for the B and green for the C). The grey band is the extrapolation to the continuum limit. Note that for $w_0 f_{\pi}$ discretization effects proportional both to $a^2$ and $a^2 m_\ell$ are visible.} 
    \label{fig:Mpi_GL_NLO_am_w0_M2b}
\end{figure}

\begin{figure}[htb!]
    \centering
    \begin{minipage}{0.5\linewidth}
    \includegraphics[width=1.0\textwidth]{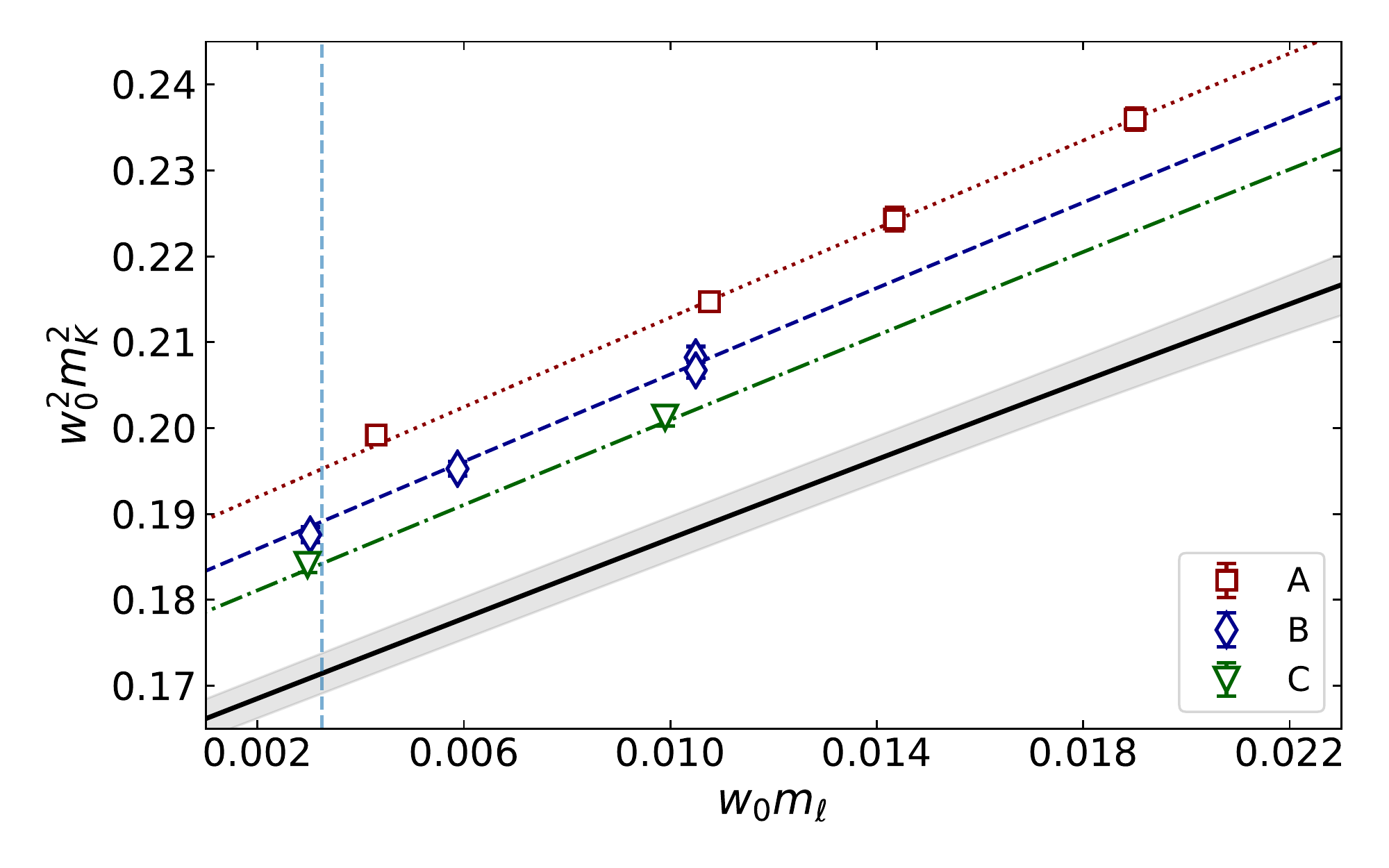}
    \end{minipage}\\
    \begin{minipage}{0.5\linewidth}
    \includegraphics[width=1.0\textwidth]{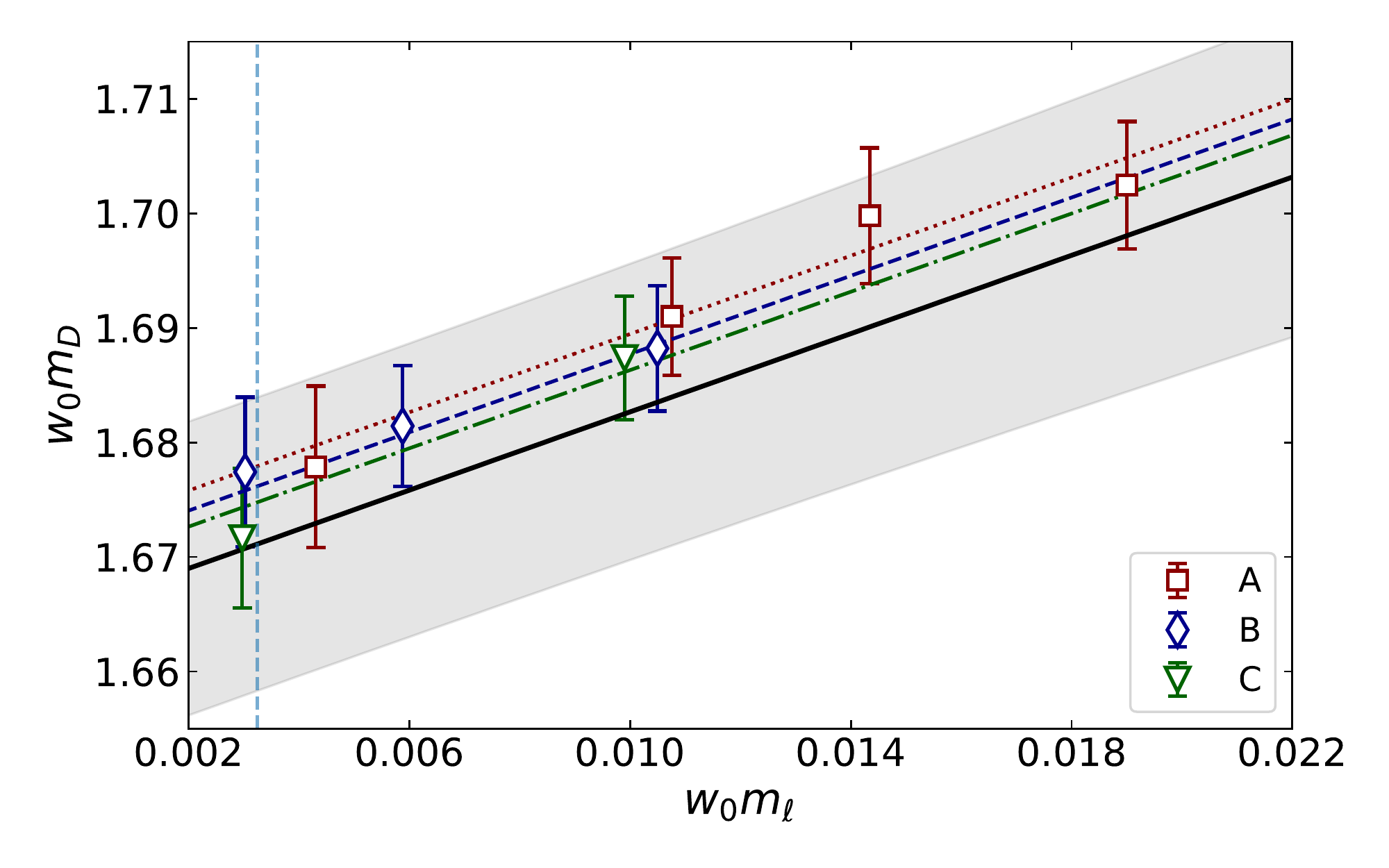}
    \end{minipage}%
    \begin{minipage}{0.5\linewidth}
    \includegraphics[width=1.0\textwidth]{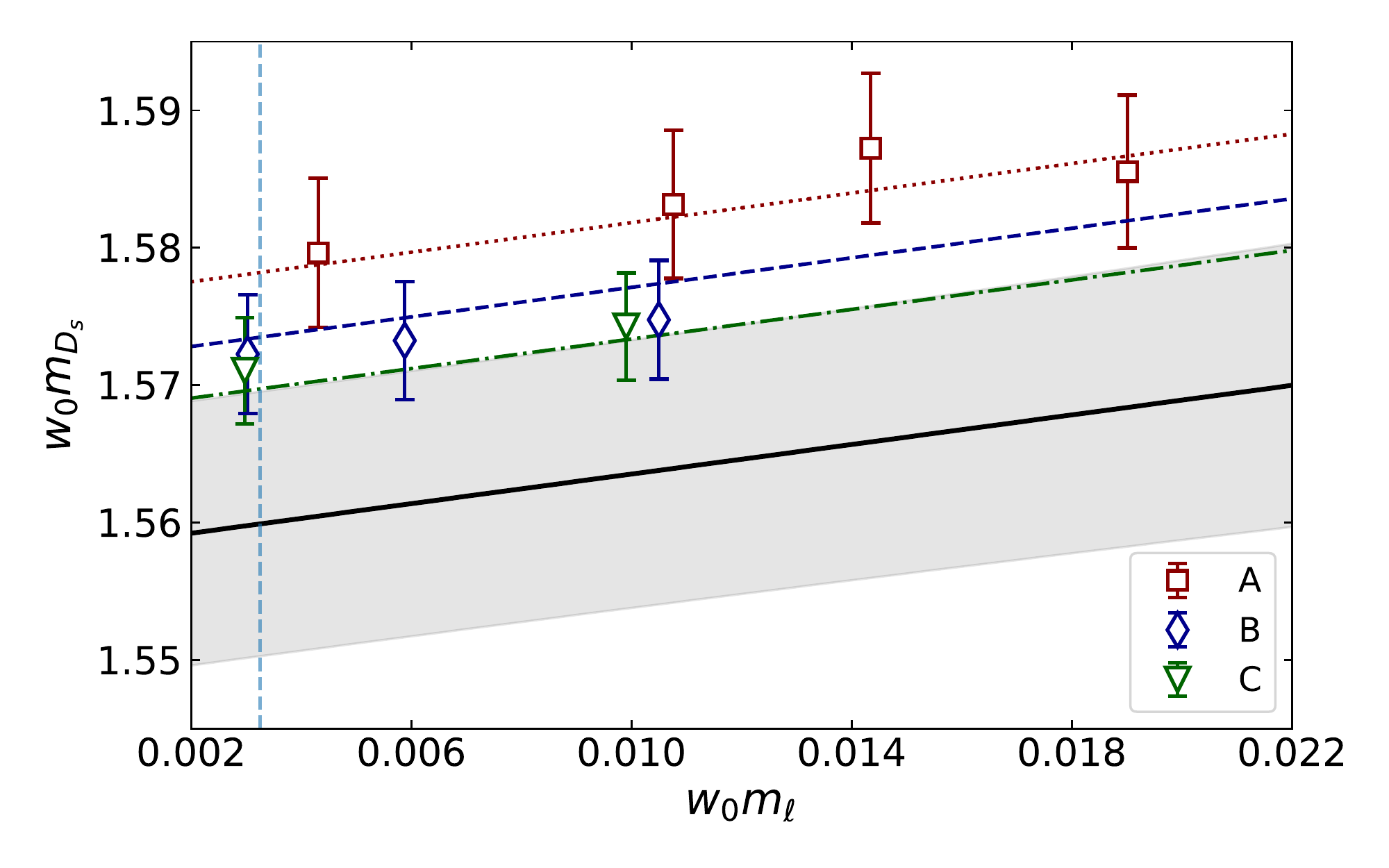}
    \end{minipage}
    \vspace{-0.5cm}
    \caption{The red, blue and green solid lines show the resulting fits using Eq.~(\ref{eq:fit_MK}) and Eq.~(\ref{eq:m_D_fit}) for the three ensembles  A, B and C, respectively. The gray line shows the continuum extrapolation, for the determination of $m_s$ (top) and $m_c$ using the mass of the D-meson (left) and the mass of the $D_s$-meson (right). }
    \label{fig:meson_ms-mc}
\end{figure}

We use the kaon mass $m_K^{\rm isoQCD}$~=~494.2(3)~MeV as input for fixing the strange quark mass. To determine the  charm quark mass, we use both  the mass of the  D- and  D$_s$-mesons, $m_D^{\rm isoQCD}=1867.0(4)$~MeV and   $m_{D_s}^{\rm isoQCD}=1969.0(4)$~MeV, respectively. We use three  reference  values for the strange and charm quark mass for all ensembles and interpolate linearly using $m_{K,D}^2 =a+b m_{s,c}s w_0$. In the case of the strange quark mass we use the  NLO ChPT inspired Ansatz~\cite{Roessl:1999iu}
\begin{equation}
    (m_K w_0 )^2 =  P_0(m_\ell w_0+m_sw_0 )\left[ 1 +P_1 m_\ell w_0+P_2 m_\ell^2 w_0^2 +  P_3  \,a^2/w_0^2\right]\,.
    \label{eq:fit_MK}
\end{equation}
In the case of the charm quark mass, given the weak dependence of $D,D_s$ meson masses on $m_\ell$, we use
\begin{equation}
 m_{D,D_s}=P_{0}^{D,D_s}+P_{1}^{D,D_s} m_\ell w_0+P_{2}^{D,D_s} a^2/w_0^2 \, . \label{eq:m_D_fit}
 \end{equation}
 The results of the chiral and continuum extrapolations are shown in Fig.~\ref{fig:meson_ms-mc}.

\section{Determination of  quark masses in the baryon sector}\label{sec:baryons}
For the determination of the light quark mass, we use the nucleon mass as an input and fit to
 the ChPT expression of Eq.~\eqref{eq:nucl_fit} to extrapolate to the physical point. To one-loop order in ChPT (up to which the nucleon mass is expanded in Eq.~\eqref{eq:nucl_fit}) we can parametrize the pion mass by $m^2_\pi = 2 B m_{ud}(1+c_2 a^2)$ obtaining the expansion
\begin{equation}
    m_N(m_{ud}) = m^0_{N} - 4 c_1 \left(2 B m_{ud}(1+c_2 a^2)\right) - \frac{3g^2_A}{16\pi f_\pi^2}\left(2 B m_{ud}(1+c_2 a^2)\right)^{3/2},
\label{eq:pion_mass_nucl}
\end{equation}
consistently with the order we are working and including $\mathcal{O}(a^2)$ effects in the pion expansion with the coefficient $c_2$. We thus have two fit parameters, $B$ and $c_2$, while the lattice spacings, $m_N^0$ and $c_1$ are determined from Eq.~\eqref{eq:nucl_fit}.

We determine the strange and charm quark masses using the experimental value of the $\Omega\,(sss)$ and $\Lambda_c\,(udc)$ masses and the lattice spacings obtained from the nucleon mass. Namely, we use $m_{\Omega}^{(phys.)}=1672.5(3)$ and $m_{\Lambda_c}^{(phys)}=2286.5(1)$ from the PDG~\cite{10.1093/ptep/ptaa104}. We parametrize the $\Omega^-$ and $\Lambda_c$ mass dependence on the strange and charm quark mass by expanding around $\tilde{m}_s$ and $\tilde{m}_c$, in the vicinity of the physical quark masses, using
\begin{align}
    \label{eq:baryonmass_vs_mus}
    m_{\Omega} = A_{\Omega} +  B_{\Omega}\, (m_s-\tilde{m}_s),\\
    m_{\Lambda_c} = A_{\Lambda_c} +  B_{\Lambda_c}\, (m_c- \tilde{m}_c) \,.
    \label{eq:baryonmass_vs_muc}
\end{align}
We employ two methods to determine $m_s$ and $m_c$: In method I  we  perform a chiral and continuum extrapolation of the $A_{\Omega,\Lambda_c}$ and $B_{\Omega,\Lambda_c}$ parameters separately using $   A_{\Omega,\Lambda_c} (a,m_{\pi}^2)   = c_1 + c_2 m_{\pi}^2 + c_3 a^2 $ and an equivalent expression for $B_{\Omega,\Lambda_c}$. In method II we adopt an iterative strategy: Namely, we start by fixing a value of the renormalized strange quark mass $m_s$ in physical units for all the ensembles and then we extrapolate to the continuum limit and physical point using the ChPT result 
\begin{equation}
m_{\Omega,\Lambda_c} = m_{\Omega,\Lambda_c}^{(0)} - 4 c_{\Omega,\Lambda_c}^{(1)} m_{\pi}^2 + d_{\Omega,\Lambda_c}^{(2)} a^2\,.
\label{eq:qmass_method2}
\end{equation}
We iterate this procedure changing the value of $m_s (m_c)$ until the resulting value of $m_{\Omega,\Lambda_c}$ given in Eq.~(\ref{eq:qmass_method2}) at the physical point and continuum limit matches the physical value $m_{\Omega}^{(phys.)}$  ($m_{\Lambda_c}^{(phys.)}$). 
\begin{figure}[htb!]
    \centering
    \begin{minipage}{0.45\linewidth}
    \includegraphics[width=1.0\textwidth]{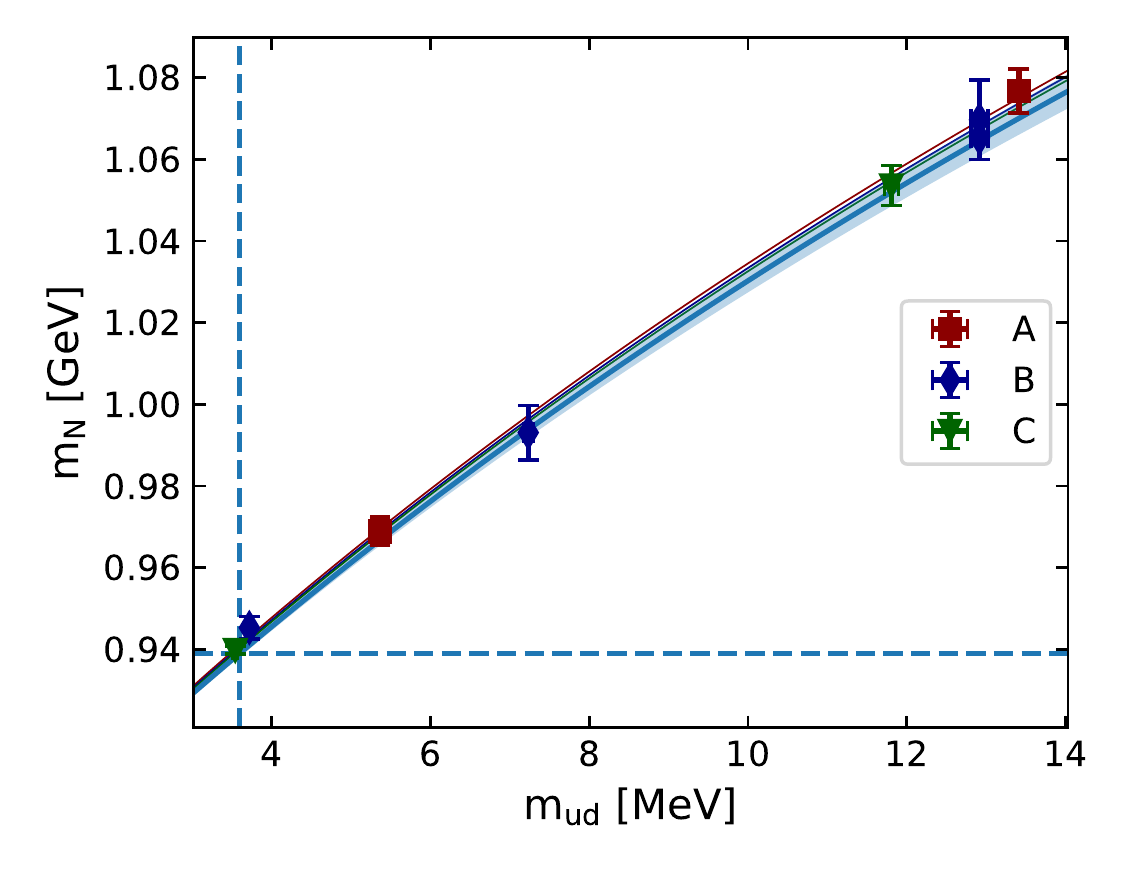}
    \end{minipage}\\
    \begin{minipage}{0.5\linewidth}
    \includegraphics[width=1.0\textwidth]{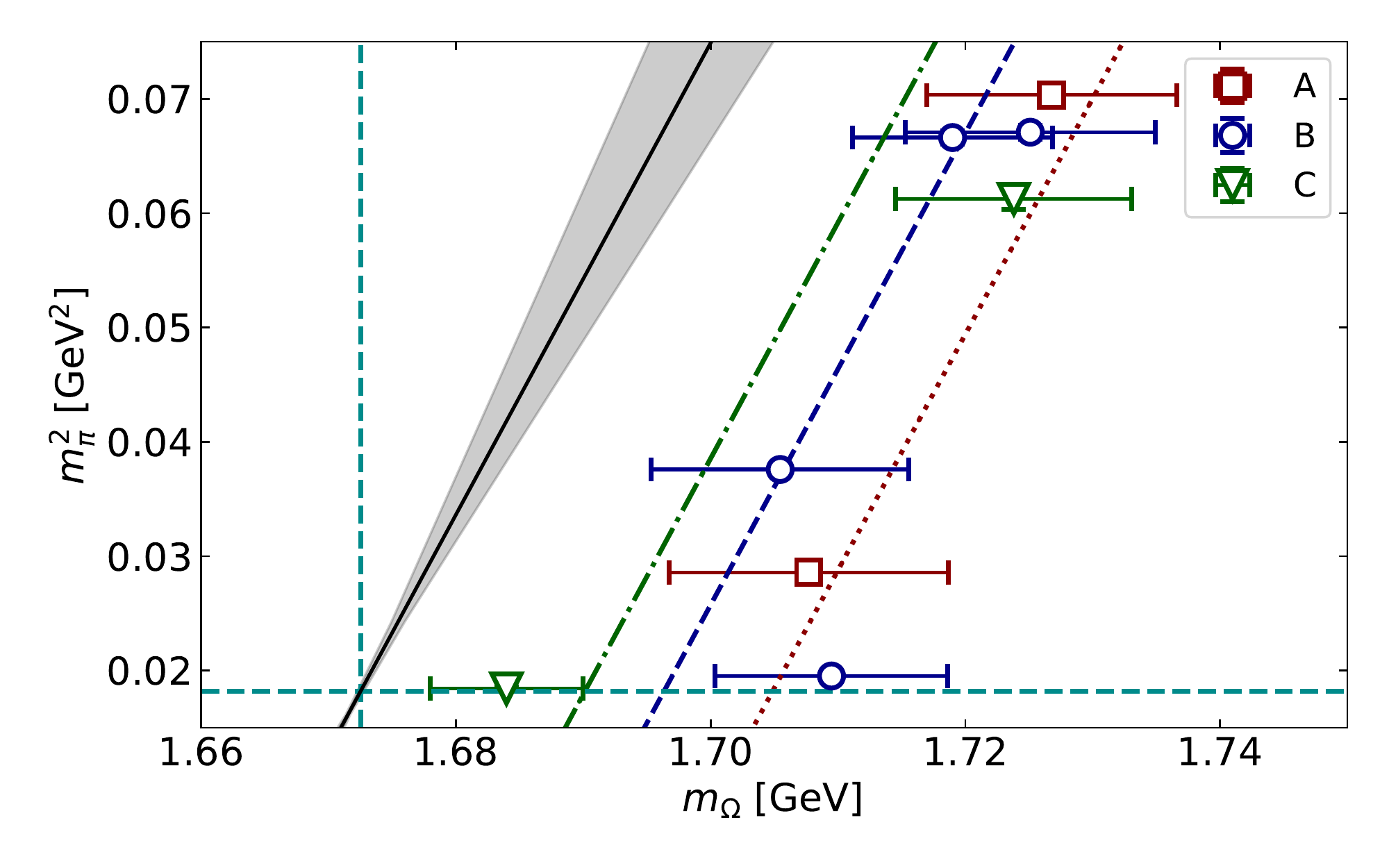}
    \end{minipage}%
    \begin{minipage}{0.5\linewidth}
    \includegraphics[width=1.0\textwidth]{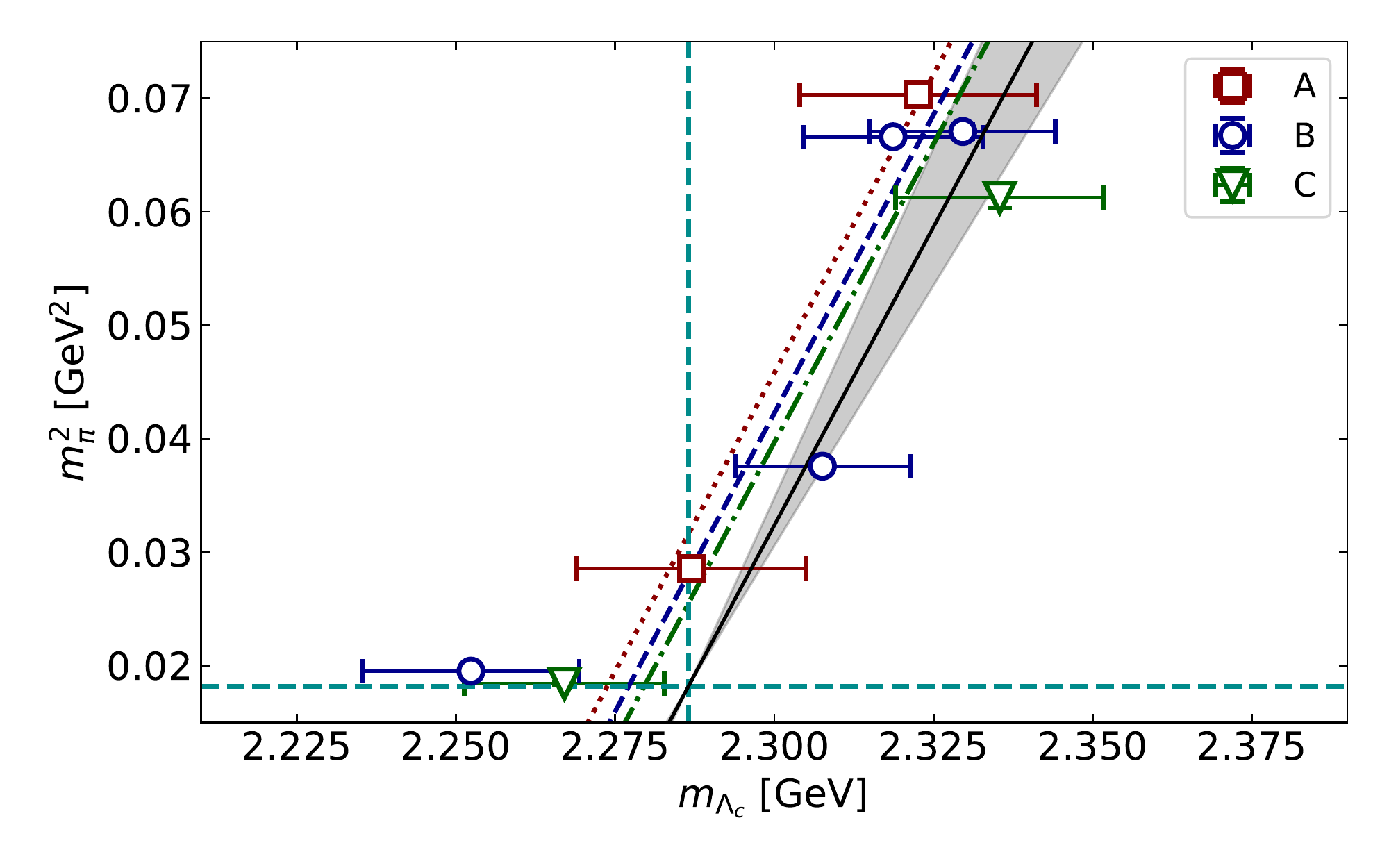}
    \end{minipage}
    \vspace{-0.5cm}
    \caption{Top: The nucleon mass $m_N$ for the A- (red), B- (blue) and C- (green) ensembles. The blue band shows the continuum extrapolation according to Eq.~(\ref{eq:pion_mass_nucl}). Bottom: We show  $m_{\Omega}$ ($\Lambda_c$) versus $m_\pi^2$, when $m_s$ ($m_c$) takes the value that reproduces the physical mass of the $\Omega$ ($\Lambda_c$) at the continuum limit as described in   method II. The dotted lines show the chiral extrapolation for the A- (red), B- (blue) and C- (green) ensembles. The solid black line shows the  continuum extrapolation using Eq.~\eqref{eq:qmass_method2} with the associated error (grey band). The horizontal and vertical dashed light blue lines represent, respectively, the physical pion and $\Omega$ ($\Lambda_c$) masses. }
    \label{fig:ms-mc}
\end{figure}
\section {Results and Conclusions}
\begin{table}[htb!]
	\small
    \centering
    \begin{tabular}{l|c|c|c||c|c}
    \hline
         & $m_{ud}$ [MeV] & $m_{s}$ [MeV] & $m_{c}$ [MeV] & $m_{s}/m_{ud}$ & $m_{c}/m_{s}$ \\
    \hline
        Meson sector &$3.689(80)(66)$ &  $101.0(1.9)(1.4)$   & $1039(15)(8)$ & $27.30(24)(14)$  & $11.43(9)(10)$ \\
        Baryon sector & $3.608(58)(^{+32}_{-19})$ & $94.9(2.4)(^{+4.1}_{-1.0})$ & $1030(21)(^{+22}_{-5})$ & $26.30(61)(^{+1.17}_{-0.33})$ & $12.04(31)(^{+58}_{-15})$ \\
        \hline
        Average  & $3.636(66)(^{+60}_{-57})$ & $98.7(2.4)(^{+4.0}_{-3.2})$ & $1036(17)(^{+15}_{-8})$ &$27.17(32)(^{+56}_{-38})$ & $11.48(12)(^{+25}_{-19})$ \\\hline\hline
         FLAG 2019 & 3.410(43) &  93.44(68) & 988(7) & 27.23(10) & 11.82(16) \\
    \end{tabular}
    \caption{The renormalized quark masses determined in the meson sector (first row) and baryon sector (second row) in the $\overline{\rm MS}$ scheme. In the third row we give the average over the values obtained in the the meson and baryon sectors, while in the last row we give the latest FLAG averages~\cite{Aoki:2019cca} for $N_f = 2+1+1$. }
    \label{tab:recap_masses}
\end{table}
In Table~\ref{tab:recap_masses} we collect the values of the quark masses obtained in Sections~\ref{sec:mesons} and \ref{sec:baryons} for the light and strange quark masses in the $\overline{\rm MS}$ scheme at 2~GeV and for the charm quark mass at 3~GeV.
\begin{figure}[htb!]
    \centering
    \includegraphics[width=1.\textwidth]{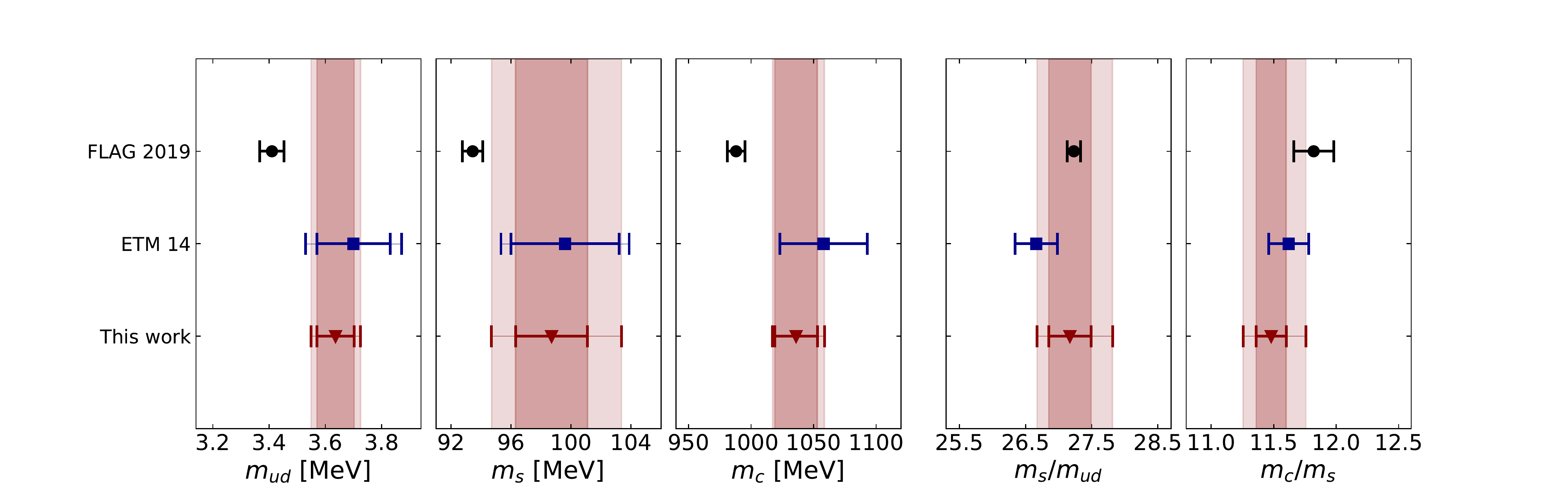}
    \vspace{-0.5cm}
    \caption{Comparison of the results average between the values determined in the meson and baryon sectors (red triangles) with the values obtained using twisted mass fermions in Ref.~\cite{Carrasco:2014cwa} (blue squares) and the $N_f=2+1+1$ averages given in the last FLAG report~\cite{Aoki:2019cca} (black circles). The shorter error bars take into account the statistical error only, while the larger represent the total error, obtained by summing in quadrature the statistical and the systematic errors.}
    \label{fig:final_res1}
\end{figure}
The final results are given in the row labeled ``Average'' of Table~\ref{tab:recap_masses} and are compared in Fig.~\ref{fig:final_res1} with those of the ETM analysis of Ref.~\cite{Carrasco:2014cwa} and the ones entering the $N_f = 2+1+1$ averages in the latest FLAG report~\cite{Aoki:2019cca}. Our results are larger by $\sim 2.5$ standard deviations in the case of $m_{ud}$ and by $\sim 2$ standard deviations in the case of $m_c$ with respect to the corresponding FLAG values. The strange quark mass tends to  also be larger, although, within the larger final error, deviates less from the FLAG result.
A very good agreement is observed for the mass ratios $m_s / m_{ud}$ and $m_c / m_s$ and the ones reported by FLAG.
\small

\section*{Acknowledgments}
We acknowledge PRACE (Partnership for Advanced Computing in Europe) for awarding us access to the high-performance computing system Marconi and Marconi100 at CINECA (Consorzio Interuniversitario per il Calcolo Automatico dell'Italia Nord-orientale) under the grants Pra17-4394, Pra20-5171 and Pra22-5171, and CINECA for providing us CPU time under the specific initiative INFN-LQCD123. We also acknowledge PRACE for awarding us access to HAWK, hosted by HLRS, Germany, under the grant with Acid 33037.
The authors gratefully acknowledge the Gauss Centre for Supercomputing e.V.~(www.gauss-centre.eu) for funding the project pr74yo by providing computing time on the GCS Supercomputer SuperMUC at Leibniz Supercomputing Centre (www.lrz.de), the projects ECY00, HCH02 and HBN28 on the GCS supercomputers JUWELS and JUWELS Booster~\cite{JUWELS} at the J\"ulich Supercomputing Centre (JSC) and time granted by the John von Neumann Institute for Computing (NIC) on the supercomputers JURECA and JURECA Booster, also at JSC. Part of the results were created within the EA program of JUWELS Booster also with the help of the JUWELS Booster Project Team (JSC, Atos, ParTec, NVIDIA). We further acknowledge computing time granted on Piz Daint at Centro Svizzero di Calcolo Scientifico (CSCS) via the project with id s702.
Part of the statistics of the cA211.30.32 ensemble used in this work was generated on the cluster at the University of Bonn, access to which the authors gratefully acknowledge.

This work has been partially supported by the Horizon 2020 research and innovation program
of the European Commission under the Marie Sk\l{}odowska-Curie grant agreement No.~765048 (STIMULATE) as well as by the DFG as a project under the Sino-German CRC110.
R.F.~acknowledges the University of Rome Tor Vergata for the support granted to the project PLNUGAMMA.
F.S.~and S.S.~are supported by the Italian Ministry of Research (MIUR) under grant PRIN 20172LNEEZ.
F.S.~is supported by INFN under GRANT73/CALAT.
P.D.~and E.F.~acknowledge support form the European Unions Horizon 2020 research and innovation programme under the Marie Sk\l{}odowska-Curie grant agreement No.~813942 (EuroPLEx). P.D.~acknowledges support from INFN under the research project INFN-QCDLAT.
M.C.~acknowledges financial support by the U.S.~Department of Energy, Office of Nuclear Physics Early Career Award under Grant No.~DE-SC0020405. 
S.B.~and J.F.~are supported by the H2020 project PRACE 6-IP (grant agreement No.~82376) and the EuroCC project (grant agreement No.~951740). F.M.~and A.T.~are supported by the European Joint Doctorate program STIMULATE grant agreement No.~765048.
K.H. is supported by the Cyprus Research and Innovation  Foundation under contract number POST-DOC/0718/0100.
F.P.~acknowledges support from project NextQCD, co-funded by the European Regional Development Fund and the Republic of Cyprus through the Research and Innovation Foundation (EXCELLENCE/0918/0129).
M.D.C.~is supported in part by UK STFC grant ST/P000630/1.

\bibliographystyle{JHEP}
{
\bibliography{refs}
}
\end{document}